\documentclass[a4paper,aps,prl,twocolumn,floats]{revtex4}
\usepackage{graphicx,dcolumn,bm}
\usepackage{amsmath,amssymb,amsfonts}
\usepackage[dvips]{hyperref}
\usepackage{subfigure}
\newcommand{\field}[1]{\mathbb{#1}}
\newcommand{\mycomm}[1]{\hfill\break
$\phantom{a}$\kern-3.5em{\tt===$>$ \bf #1}\hfill\break}
\newcommand{\mycommA}[1]{\hfill\break
$\phantom{a}$\kern-3.5em{\tt***$>$ \bf #1}\hfill\break}

\catcode`\@=11 % This allows us to modify PLAIN macros.
\def\lsim{\mathrel{\mathpalette\@versim<}}
\def\gsim{\mathrel{\mathpalette\@versim>}}
\def\@versim#1#2{\vcenter{\offinterlineskip
        \ialign{$\m@th#1\hfil##\hfil$\crcr#2\crcr\sim\crcr } }}
\catcode`\@=12 % at signs are no longer letters
\def\beq{\begin{equation}}
\def\eeq{\end{equation}}

\def\FP{\hbox{\tiny FP}}

\def\mysim{\kern -.1667em\lower0.8ex\hbox{$\tilde{\phantom{a}}$}}

\begin{document}

%\title{\Large Looking in through the conformal window}
%\title{\Large Squinting through the conformal window}
\title{\Large Delineating the conformal window}
\author{Mads T. Frandsen}
\email{m.frandsen1@physics.ox.ac.uk}
\author{Thomas Pickup}
\email{t.pickup1@physics.ox.ac.uk}
\author{Michael Teper}
\email{m.teper1@physics.ox.ac.uk}
\affiliation{Rudolf Peierls Centre for Theoretical Physics, University of Oxford, Oxford OX1 3NP, United Kingdom}
%\date{\today}
\begin{abstract}
We identify and characterise the conformal window in gauge theories relevant 
for beyond the standard model building, e.g. Technicolour, using the criteria of 
metric confinement and causal analytic couplings, which are known to be consistent with the 
phase diagram of supersymmetric QCD from Seiberg duality. Using these criteria we 
find perturbation theory to be consistent throughout the predicted conformal window 
for several of these gauge theories and we discuss recent lattice results in the light 
of our findings. 
\end{abstract} 
\maketitle

\section{The conformal window}

In a generic non-Abelian gauge theory with gauge group $G$ and $N_f$ fermions 
transforming according to a representation $R$ of $G$ we expect there to be a 
conformal window \cite{BZ}, i.e. a region $N_f^{\rm{II}}<N_f<N_f^{\rm{I}}$ 
for which the theory is asymptotically free at short distances while the 
long distance physics is scale-invariant and typically governed by a 
non-trivial fixed-point. In this paper we consider such theories with fermions in a single representation of the gauge groups $SU, SO, Sp$.

The upper boundary of the conformal window is determined in perturbation 
theory from the $\beta$ function: 
\beq
\beta(x)\,\equiv\, \frac{dx}{d\ln(Q^2)}\,=\,-
\left(\beta_0x^2+\beta_1x^3+\cdots\right), 
\label{betafunction}
\eeq
at a small value of the coupling $x\equiv\alpha_s/\pi$. The first two coefficients of the expansion
\cite{oneloop,twoloops} are universal and independent of the renormalisation 
group scheme:
\begin{eqnarray}
4 \beta_0 &=& \frac{11}{3}C_2(G)- \frac{4}{3}T(R)N_f \label{beta0}\\
16 \beta_1 &=&  \frac{34}{3} C_2^2(G)  
 - \frac{20}{3}C_2(G)T(R) N_f   
\nonumber \\
       &\phantom{-}& - \, 4C_2(R) T(R) N_f     .
\label{beta1}
\end{eqnarray}
When $\beta_0$ changes sign, from 
positive to negative at 
\begin{eqnarray}
N_f^{\rm{I}} = \frac{11}{4} \frac{C_2(G)}{T(R)} \ ,
\label{NfI}
\end{eqnarray}
the theory changes from the asymptotically free conformal phase to the 
infrared free phase. This is the upper boundary of the conformal window, 
coinciding with the loss of asymptotic freedom (LOAF), and the transition 
point in $N_c=3$ QCD is at $N_f^{\rm{I}}=16.5$.  For $N_f$ just below 
this upper boundary, Eqs.(\ref{beta1},\ref{NfI}) imply that $\beta_1 < 0$, 
and so $\beta(x)$ will have a non-trivial zero at 
\hbox{$x_{\FP}\simeq-\beta_0/\beta_1>0$}. 
The fixed point coupling $x_{\FP}$ approaches zero as $N_f$ approaches $N_f^{\rm{I}}$ 
from below. The smallness of $x_{\FP}$ just below $N_f^{\rm{I}}$ justifies the use of the 2-loop $\beta$ 
function. Thus the transition to the infrared free phase is always via a 
conformal phase \cite{BZ} and this is independent of the fermion representation.

The lower boundary of the conformal window, $N_f^{\rm{II}}$, below which 
confinement and chiral symmetry breaking typically set in, is much harder to 
determine. From the two-loop $\beta$-function, the fixed point is lost and the lower boundary of 
the conformal window would be reached from above when $\beta_1=0$. However, 
this not only ignores higher order corrections but also neglects 
non-perturbative effects which, generally, are expected to become important 
towards the lower end of the conformal window, where the 2-loop
estimate of the fixed point coupling is becoming large, $x_{\FP}\gtrsim 1$.

While the lower boundary of the conformal window is of theoretical interest 
in its own right, its current importance arises from its central role in technicolour 
models \cite{Weinberg:1979bn} with walking dynamics \cite{Holdom:1981rm,Lane:1989ej} and, 
in particular, of more recent models such as minimal walking 
technicolour \cite{Sannino:2004qp} and conformal technicolour 
\cite{Luty:2004ye}.
Therefore, a lot of effort has recently gone into exploring this region, 
using both lattice \cite{Catterall:2007yx,Appelquist:2007hu,Hietanen:2008mr,Shamir:2008pb,DelDebbio:2008zf,Deuzeman:2009mh,
Fodor:2009ar,Yamada:2009nt,Bursa:2009we,DeGrand:2009hu,Jin:2009mc,Yamada:2010wd,Kogut:2010cz,Hasenfratz:2010fi,DelDebbio:2010hx,DeGrand:2010na} and approximate 
analytical \cite{Appelquist:1988yc,Gardi:1998ch,Appelquist:1999hr,Dietrich:2006cm,Ryttov:2007sr,Ryttov:2007cx,DelDebbio:2008wb,Sannino:2008pz,
Poppitz:2009uq,Armoni:2009jn,Sannino:2009me,Dietrich:2009ns,Yamawaki:1996vr} methods. In principle the 
former should provide a definitive answer: however, it has become clear, 
from the pioneering lattice calculations, that identifying and 
characterising (near-)conformal theories on a lattice is a very challenging 
problem. So it remains important to try and gain as much analytical insight 
as possible.

Since it is the chiral symmetry breaking of technicolour that drives
the interesting `walking' scenarios, it is natural to look to analytic
methods that estimate its onset. The standard technique involves the
use of Schwinger-Dyson (SD) equations in a ladder-like approximation \cite{Appelquist:1988yc,Dietrich:2006cm,Ryttov:2007sr}. While 
this does make a prediction for the value of $N_f$ at which chiral
symmetry is spontaneously broken, the credibility of the estimate
is called into question by the fact that in the case of $\mathcal{N}=1$ supersymmetric QCD (SQCD), where Seiberg
duality \cite{Seiberg:1994pq} allows us to calculate  the value of $N_f^{\rm{II}}$ exactly,
the SD estimate is far above the known value
\cite{Appelquist:1997gq}. Thus it is useful to look for other
analytical estimates which can help determine where conformality may be lost.

Here we wish to discuss two such methods, both of which have been
extensively discussed in the 1990's in related and overlapping 
contexts. First we shall discuss the criterion of `metric confinement' 
\cite{Oehme_metric_confinement}, which provides a {\it lower bound} 
on the value of $N_f$ at which confinement occurs and thus also 
for the value of $N_f^{\rm{II}}$ at which
conformality is lost. Secondly we discuss the range of validity of 
perturbation theory within the conformal window following 
\cite{Gardi:1998ch,Gardi:1998rf,Gardi:1998qr} and we compare our findings with 
lattice simulations of these theories. 

\subsection{Metric confinement}

Metric confinement determines when transverse gluons are not 
part of the physical Hilbert space from the properties of the transverse 
gluon propagator, $D(Q^2,\mu^2,g)$, where $\mu^2$ is the renormalisation 
scale. We refer the reader to \cite{Oehme_metric_confinement} for a detailed exposition of metric confinement. The condition can be formulated 
(working always in Landau gauge) in terms of a superconvergence relation for the absorptive 
part $\rho(k^2,\mu^2,g)=(1/\pi)\,{\rm Im}\left\{D(-k^2,\mu^2,g)\right\}$  
of the gluon propagator \cite{Oehme_metric_confinement}:
\beq
\int_{0^-}^{\infty}dk^2\,\rho(k^2,\mu^2,g)\,=\,0.
\label{superconvergence_relation}
\eeq
Because of the known analyticity properties of the propagator $D$, 
Eq.~(\ref{superconvergence_relation}) is equivalent to the vanishing 
of the integral of $D$ around the contour at $|k^2|=\infty$ \cite{Oehme_metric_confinement}. Thus,
if $D(Q^2,\mu^2,g)$ vanishes fast enough as $|Q^2| \to \infty$, one will 
indeed have metric confinement. Asymptotic freedom then allows us to 
determine whether it does so or not from the value of the appropriate
anomalous dimension. The condition for metric confinement, in terms
of the 1-loop anomalous dimension of the gluon propagator $\gamma_{00}$ 
can be seen to be \cite{Oehme_metric_confinement}: 
\beq
\gamma_{00}= -\frac 14\left( \frac{13}{6}C_2(G)-\frac 43T(R)N_f\right) < 0 .
\label{gamma00}
\eeq
Note that because we are interested in the value of $D$ as $|Q^2| \to \infty$,
the 1-loop perturbative value of $\gamma_{00}$ is exact for our purposes:
when Eq.~(\ref{gamma00}) holds the theory confines and conformality has been
lost. Metric confinement is claimed to provide a sufficient but not 
necessary condition for confinement and therefore Eq.~(\ref{gamma00})
provides a lower bound on the lower boundary of the conformal window:
\beq
N_f^{\rm{II}} \geq N_f^{\rm{MC}} \equiv 13C_2(G)/8 T(R) \ .
\label{MCbound}
\eeq
We also note from Eq.~(\ref{NfI}) that this bound is strictly less than
the upper edge of the conformal window: $N_f^{\rm{MC}}<N_f^{\rm{I}}$.
So metric confinement always leaves a finite window of opportunity
for conformality.

This lower bound on $N_f^{\rm{II}}$ 
\cite{Oehme_metric_confinement}
is plotted for $SU$ and $SO$ gauge theories with fermions in single- 
and two-index representations, as the thick dotted line, in 
Figs.~\ref{fig12} and~\ref{fig22}. We discuss the implications
later in the paper.

Just as with the SD estimates, it is useful to test this
bound in SQCD. Remarkably, one 
finds that the 
lower bound on $N_f^{\rm{II}}$ from metric confinement {\it coincides} 
with the value of $N_f^{\rm{II}}$ that is determined from Seiberg duality 
\cite{Seiberg:1994pq}. This has been shown for both $SU$ and $SO$ gauge
groups \cite{Oehme:1995ae,Oehme:1997ky,Gardi:1998ch} and is also 
the case for $Sp$ gauge groups, as we have checked ourselves.
Such agreement is particularly significant in the case of SQCD as
it is known \cite{Seiberg:1994pq} that here the loss of
conformality is through the onset of confinement and not of chiral
symmetry breaking -- the latter occurring at a much smaller value 
of $N_f$. (This provides a striking counterexample to the earlier wisdom
that confinement necessarily entails chiral symmetry breaking.)

It is also interesting to consider supersymmetric Yang Mills with fermionic matter in higher representations where there is no known Seiberg dual. In these cases if one determines the lower boundary of the conformal window using the Novikov-Shifman-Vainshtein-Zakharov (NSVZ) beta function 
for supersymmetric theories \cite{Novikov:1983uc} by setting $\gamma=1$ (the unitary bound in these theories) \cite{Ryttov:2007sr}, which in the case of SQCD is known to reproduce the result from Seiberg duality, we find that even in these theories metric confinement coincides with this result.

Motivated by these examples, we shall assume in the remainder of this paper
that metric confinement is (usually) not just a sufficient but also  
a necessary condition for confinement to occur.

\subsection{Perturbation theory and analyticity}

At large momentum transfer $Q^2$, the coupling constant behaves as 
$x(Q^2)\sim \frac{1}{\beta_0 \ln \left( Q^2/\Lambda^2\right) }$.
At 1-loop this simple expression is valid for all $Q^2$, so that $x(Q^2)$
diverges at  $Q^2 = \Lambda^2$. Thus if we attempt to calculate some physical
quantity in a convergent power series in the 1-loop running coupling, 
this physical quantity will inherit this Landau singularity. 
This, however, will in general violate the known analyticity properties
of such a physical quantity, which typically involves specific poles
and cuts corresponding to asymptotic states. Thus we see that
perturbation theory in the 1-loop running coupling cannot be adequate
and that this is immediately visible from the unphysical analytic structure 
of the coupling. This suggests that, more generally, the analytic structure of a running
coupling can indicate whether there is any possibility of perturbation
theory providing a complete description of the physics.

Here we are interested in studying the conformal window and, in this
case, we have an infra-red fixed point, so the coupling is bounded
by $0\leq x(Q^2) \leq x_{FP}$ for $0 \leq Q^2 < \infty$ and so cannot 
have such a divergence. In particular this is the case if we use the 
2-loop coupling and if $\beta_1 < 0$. 
%In the conformal window we expect the infrared phase to be a Coulomb phase of massless particles. 
As we approach the upper bound, $N_f \to N_f^I$, the coupling becomes
weak on all scales and we may expect perturbation theory to work well.
In that case, the coupling $x(Q^2)$ should manifest the analytic 
structure of a typical physical quantity i.e.  a cut for 
$k^2 = -Q^2 \geq 0$ corresponding to the production of massless
particles, and no other unphysical singularities in the entire complex
$Q^2$ plane. If this is so then it is said to be {\em causal analytic} 
and indeed this turns out to be the case 
for  $N_f \to N_f^I$ \cite{Gardi:1998rf}. If we now decrease $N_f$ away 
from $N_f^I$ then, as long as the coupling remains causal analytic,
it is consistent for the physics to be perturbative. 
As we continue decreasing $N_f$, at some point $x(Q^2)$ will acquire unphysical singularities
in the complex $Q^2$ plane. These might be poles
or cuts. At this point the coupling ceases to be causal analytic
and signals the fact that there must now be non-perturbative
contributions that will serve to restore the correct analytic 
structure to the quantity being calculated. These may lead to confinement and/or chiral symmetry breaking and hence the
loss of conformality.

%and turns 
%out to give a reasonably small coupling in the conformal window.

The two loop $\beta$-function can be integrated explicitly in
terms of the Lambert W-function \cite{Lambert} defined by 
$W(z) \exp\left[W(z)\right]=z$, giving \cite{Gardi:1998ch,Gardi:1998rf,Gardi:1998qr}
\beq
\begin{array}{c}
\displaystyle
x(Q^2)=-\frac{1}{c}\,\,\frac{1}{1+W(z)} \ , 
\quad c=\frac{\beta_1}{\beta_0} \ , \nonumber\\
\phantom{a}\\
\displaystyle
z = 
%-\frac{4}{c}\exp\left(-1-\beta_0 t/c\right)
%=
-\frac{1}{c\, e}
\left(\frac{Q^2}{\Lambda^2}\right)^{-\beta_0/c} \ .
\end{array}
 \label{W_sol_2loop} 
\eeq
While $W(z)$ is a multi-valued function with an infinite number of branches, 
the unique branch for $c<0$ with a real coupling along the positive real 
$Q^2$ axis is the principal branch denoted $W_0(z)$ 
\cite{Gardi:1998ch,Gardi:1998rf,Gardi:1998qr}. 
The requirement for this coupling to be causal translates into the criterion
\beq
0<-\beta_0^2/\beta_1<1 
\label{gamma_first_def}
\eeq
Note that as one approaches the upper bound to the conformal window,
$\beta_0 \to 0^+$ while $\beta_1 < 0$, this bound is always satisfied,
i.e. the coupling is causal analytic in this Bank-Zaks limit, 
as one might expect.
Note also that this is a stronger criterion than just requiring that the 
two-loop $\beta$-function have a fixed-point since, as $\beta_1 \to 0^-$ one violates the bound in Eq.~\ref{gamma_first_def}. Reflecting this, the analytically 
continued coupling will acquire singularities in the complex plane at a larger value of $N_f$ than where  
the Landau singularity appears
\cite{Gardi:1998ch,Gardi:1998rf,Gardi:1998qr}.

We observe from Eq.~\ref{gamma_first_def} that the coupling is causal analytic all the way down to 
$N_f^{\rm{MC}}$ provided $C_2(R)>\frac{11}{26}C_2(G)$, 
which is true in all cases, except for $SU(2)$ (and $Sp(4)$) with fundamental fermions. 
Hence it is also the case all the way down to $N_f^{\rm{II}}$ if we 
accept the bound in Eq.~(\ref{MCbound}).
For multi-flavor QCD this was already noted in \cite{Gardi:1998ch}. 
This demonstrates that while causal analyticity may be a necessary
condition for non-perturbative physics to be unimportant, it is not 
sufficient. In \cite{Gardi:1998ch} it was also shown that in SQCD 
(whose $\beta$-function differs from Eqs.~(\ref{beta0},\ref{beta1})
because of the presence of scalars and gluinos) analyticity breaks down {\it before} 
$N_f^{\rm{II}}$ is reached. This fits in with the requirements of the
weak-strong coupling Seiberg duality \cite{Seiberg:1994pq} where the
lower and upper boundaries of the conformal windows of the dual theories
are mapped into each other, which implies that near the lower
boundary the theory must be strongly coupled. This demonstrates 
that when analyticity breaks down, so that non-perturbative 
physics must be present, this does not necessarily entail confinement,
chiral symmetry breaking, or indeed the loss of conformality.

The analyticity bound in Eq.~(\ref{gamma_first_def}) is obtained from
the 2-loop $\beta$-function and so can only be regarded as approximate.
(Although in \cite{Gardi:1998qr} it was shown that going to 3-loops,
utilising a particular Pad\'e approximant functional form, does 
not alter the conclusions, as long as the 3-loop coefficient of the 
$\beta$-function is not very large.) Moreover, we expect that the 
perturbative expansion for $\beta(x)$ cannot be better than  asymptotic, 
with corrections $\sim \exp\{-c/x\}$ that mimic non-perturbative
contributions. Roughly speaking, we would expect the causal analyticity
calculated at 2-loops to be reliable as long as the coupling $x(Q^2)$
is not too large anywhere in the complex $Q^2$ plane. 

When judging whether a coupling is 'small' or 'large' it is in some sense more natural to use the scaled ('t Hooft) coupling $N_c x$ instead of $x$ as, at large $N_c$, $x\sim N_c^{-1}$ while the $n$-th coefficient of the $\beta$-function scales as $\beta_n \sim N_c^{n+1}$, and similarly for the anomalous dimension. As an example, the mass anomalous dimension of an adjoint fermions is given by $\gamma_{\rm Adj}=\frac{3}{2} (N_c x)+O(N_c^2 x^2)$. We shall therefore
calculate $\max_{Q^2 \in \field{C}} |N_c x(Q^2)|$ using the correct analytic
continuation of $x$ from the 2-loop $\beta$-function and use the magnitude
of the result as a supplementary criterion for judging the reliability 
of any argument from analyticity.

For the moment we simply plot the value of $N_f$ where analyticity is lost, 
and hence where perturbation theory signals its own breakdown according to 
the criterion in  Eq.~(\ref{gamma_first_def}), as the black solid lines in 
Figs.~\ref{fig12} and ~\ref{fig22}. We interpret these results below.

\subsubsection{Analyticity with the all-orders beta-function conjecture}
Inspired by the NSVZ beta function \cite{Novikov:1983uc}, 
an all-orders (AO) beta function for $SU(N)$ gauge theories with 
any matter representation was conjectured in \cite{Ryttov:2007cx} and further studied in \cite{Dietrich:2009ns}. It reads:
\begin{eqnarray}
\label{AOBF}
\beta(x) &=&- \beta_0 x^2 \frac{1-  T(R)\,N_{f}\,\gamma(x)/(6\beta_0)}{1- \frac{x}{2} C_2(G) \left( 1+ \frac{2\beta_0'}{\beta_0} \right)x} \ ,
\end{eqnarray}
where, 
\begin{eqnarray}
\gamma(x)=\frac{3}{2}C_2(R)x +O(x^2) \ , \quad 4 \beta_0' =  C_2(G) - T(R)N_f \ .
\end{eqnarray}
Here, $\gamma\equiv -\frac{d\ln m}{d\ln \mu}$ is the fermion mass anomalous dimension, and solving for $\gamma$ at a fixed point, i.e $\beta = 0$, 
yields $\gamma = \frac{11C_2(G)-4T(R)N_f}{2T(R)N_f}$ which increases as 
$N_f$ is decreased. Since $\gamma\leq 2$ is a rigorous bound from unitarity \cite{Mack:1975je}, this provides 
a different {\it lower bound} on $N_f^{\rm{II}}$,
\beq
N_f^{\rm{AO}}=\frac{11}{8}\frac{C_2(G)}{T(R)}
\label{NfII_AO}
\eeq
which we see is slightly below the bound provided by metric confinement in
Eq.~(\ref{MCbound}). 

\begin{figure*}[htp!]
\centering
\mbox{\subfigure{\includegraphics[width=2.25in]{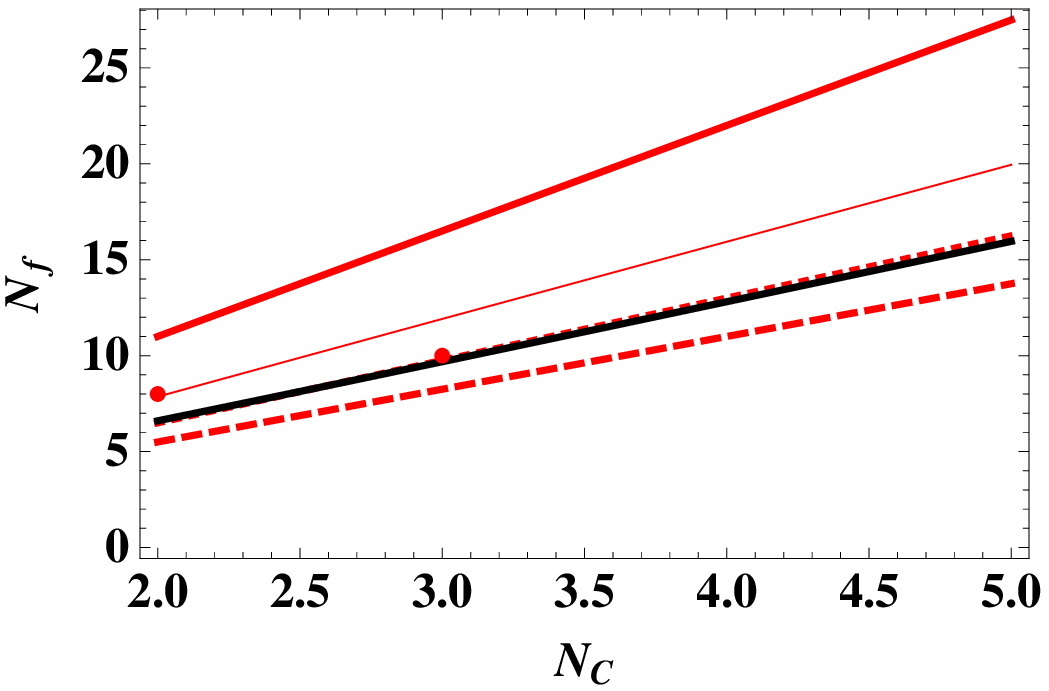}}\quad
\subfigure{\includegraphics[width=2.25in]{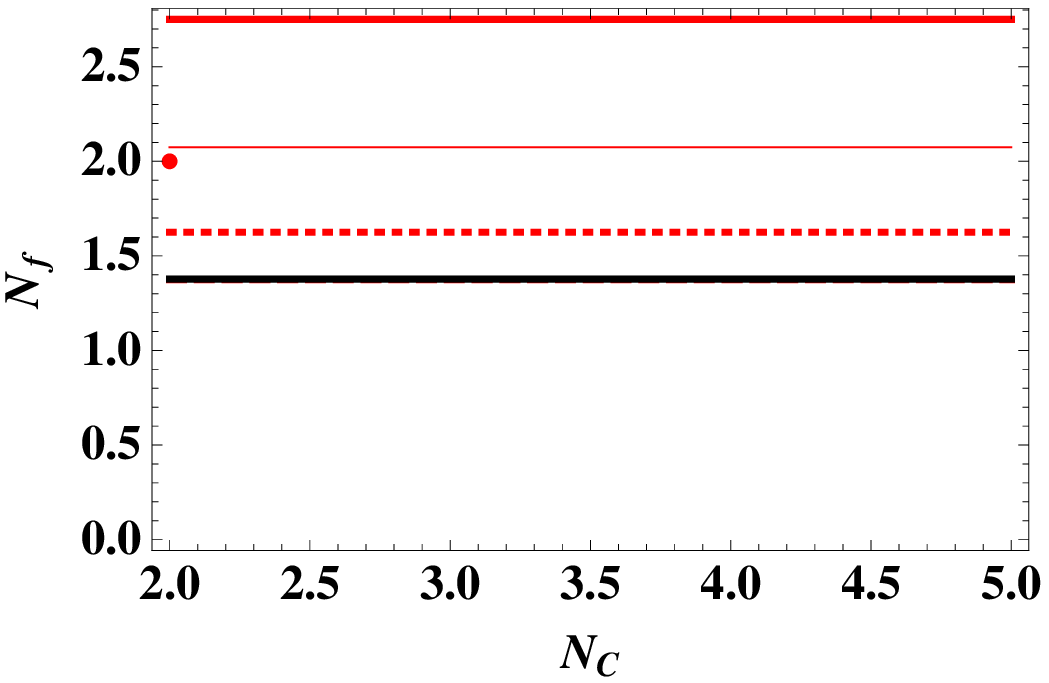}}\quad
\subfigure{\includegraphics[width=2.25in]{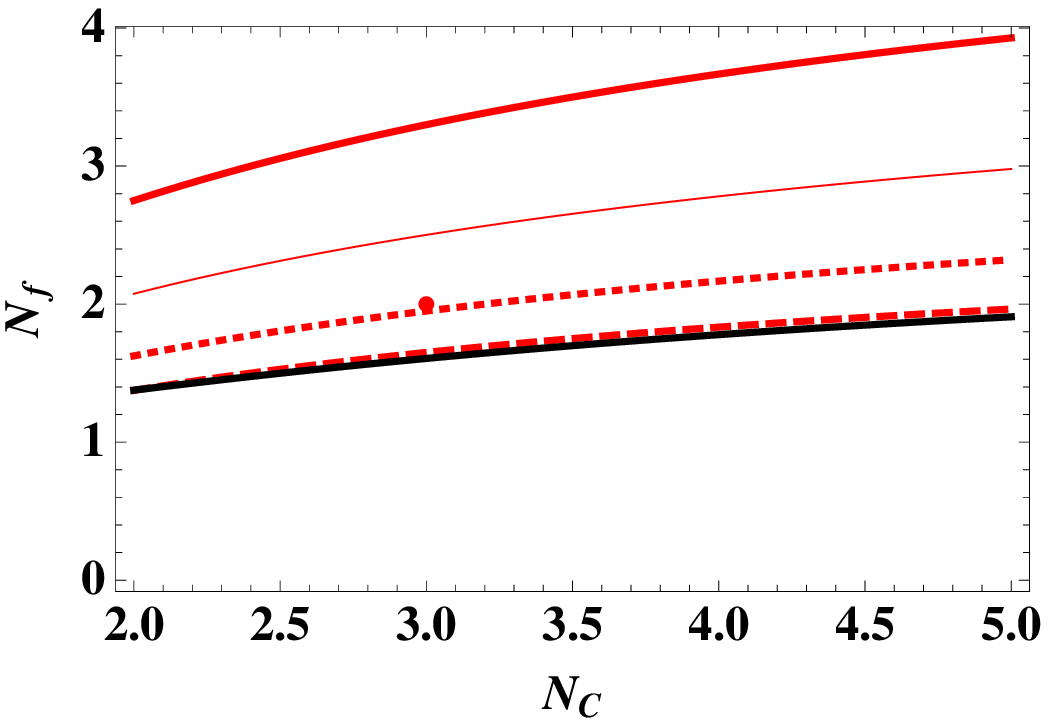}}}
\caption{Conformal windows for $SU$ theories with Dirac fermions in the fundamental (left), adjoint (mid) and two-index symmetric (right) representations. On all three figures the curves indicate $N_f^I$ 
(thick upper solid) and $N_f^{II}$ according to SD (thin solid), metric (thick dotted), AO $\beta$-function with 
$\gamma=2$ (thick dashed) and finally loss of causal analyticity (thick lower solid, black). For the adjoint representation the latter two very nearly coincide. The theories discussed in the main text are indicated with red dots.} 
\label{fig12}
\end{figure*}

\begin{figure*}[htp!]
\centering
\mbox{\subfigure{\includegraphics[width=2.2in]{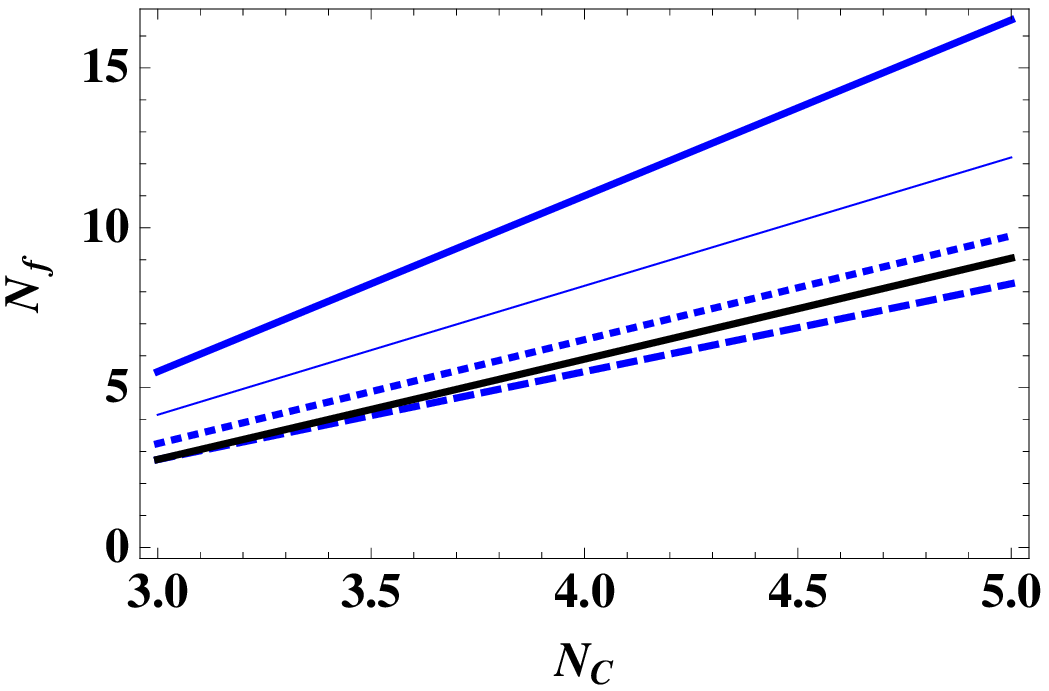}}\quad
\subfigure{\includegraphics[width=2.2in]{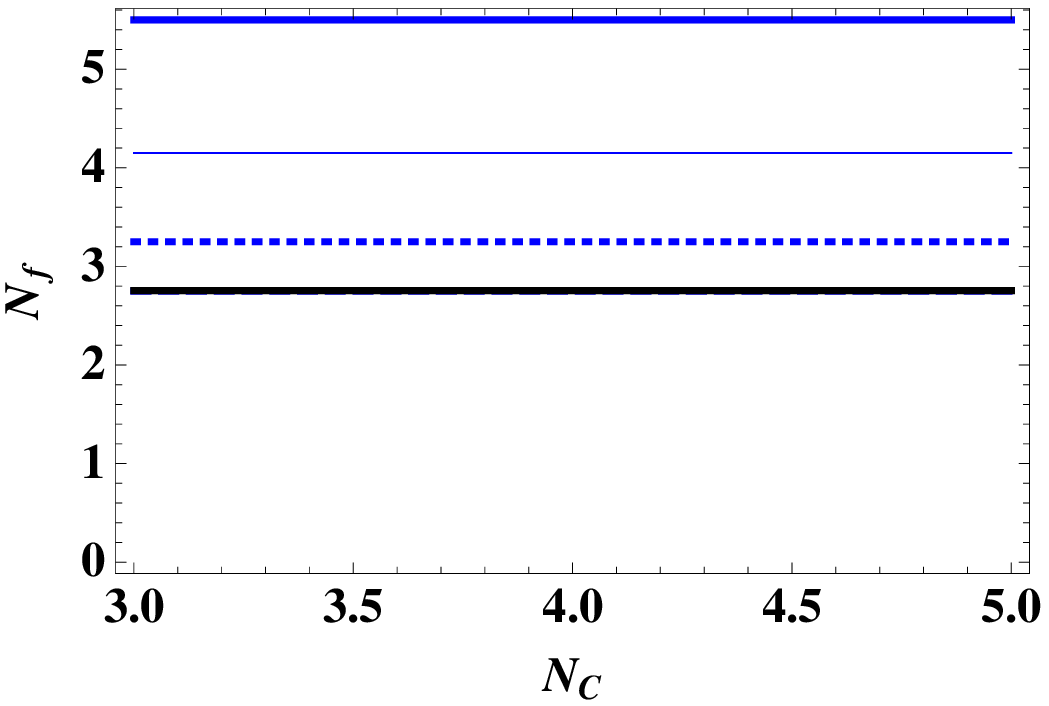}}\quad
\subfigure{\includegraphics[width=2.2in]{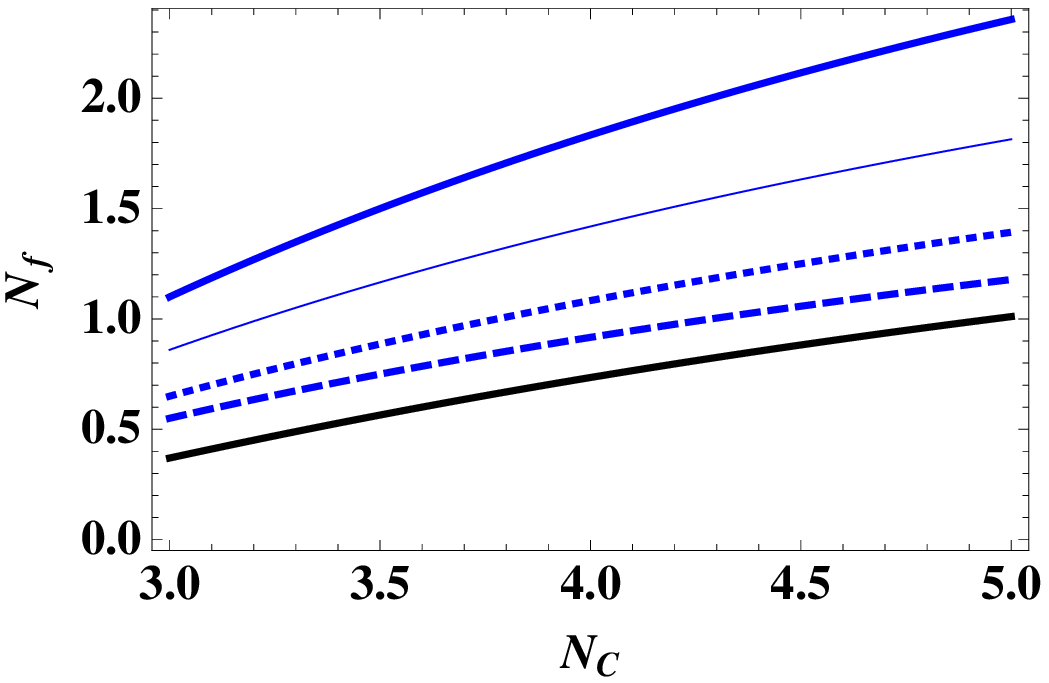}}}
\caption{Same as Fig.~\ref{fig12} but for $SO$ theories.} 
\label{fig22}
\end{figure*}

In Figs.~\ref{fig12} and~\ref{fig22} we plot this lower bound,
$N_f^{\rm{AO}}$, as a thick dashed line. For the adjoint 
representations this line is invisible because it exactly coincides with 
the thick solid line that represents the loss of causality in the two-loop 
$\beta$-function.

%\begin{figure*}[htp!]
%\centering
%\mbox{\subfigure{\includegraphics[width=3in]{confwindowSUSYM.eps}}\quad
%\subfigure{\includegraphics[width=3in]{confwindowSOSYM.eps}}}
%\caption{Conformal windows for SU (left, red), SO(right, blue) with fermions 
%in the 2-index symmetric rep. The dots indicate to the $SU(2)$ and $SU(3)$ 
%theories with 2 flavors of sextet fermions discussed in the text} 
%\label{fig32}
%\end{figure*}

We observe that if we restrict the matter anomalous dimension $\gamma$ 
to first order in $x$ then this all orders $\beta$-function may be 
integrated exactly, yielding: 
\beq
\begin{array}{c}
\displaystyle
x(Q^2)=\frac{1}{E_1}\,\,\frac{1}{1+G_1 W(z)} \ , 
\quad G_1\equiv 1-\frac{D}{E_1} \ ,\nonumber\\
\phantom{a}\\
\displaystyle
z = \frac{1}{G_1}\exp\left(-1/G_1\right) \left(\frac{Q^2}{\Lambda^2}\right)^{\frac{\beta_0}{E_1 G_1}} \ ,
\end{array}
 \label{W_sol_NSVZ}
\eeq
where 
%\begin{align}
%G\equiv 1-\frac{D}{E_1} \ , \quad B=\frac{\beta_0}{2} \nonumber
%\end{align}
%and
\beq
E_1=C_2(r\,)T(R)\,N_{f}/(4\beta_0) \ , \quad D=\frac{1}{2} 
C_2(G)\left( 1+ \frac{2\beta_0'}{\beta_0} \right) \ . \nonumber
\eeq
We can integrate the AO $\beta$-function in this approximation of $\gamma$ as it has the same structure as a Pad\'e approximant to the 3-loop $\beta$-function which is integrable in terms of the $W$-function \cite{Gardi:1998qr}.
The condition for having a causal coupling thus becomes $\beta_0 < E_1-D$ which 
is identical to the criterion for the two loop coupling being causal. 

Similarly the coupling is causal analytic all the way down to 
$N_f^{\rm{AO}}$ provided $C_2(R)>\frac{199}{198} C_2(G)$, which for the theories considered here, is generally only the case for the two-index symmetric representation.

\subsection{Comparing with lattice data and other methods}

Both the criterion of metric confinement and that of causal analyticity 
are consistent with the properties of the conformal window in SQCD as predicted from 
Seiberg duality. It is therefore interesting to ask what these criteria 
predict for the non-supersymmetric theories that are being
investigated using lattice techniques. These theories include SU(2)
and SU(3) with a `large' number of fundamental (F) fermions \cite{Appelquist:2007hu,Deuzeman:2009mh,Yamada:2009nt,Jin:2009mc,Yamada:2010wd,Hasenfratz:2010fi}, SU(2) with
2 adjoint (Adj) fermions \cite{Catterall:2007yx,DelDebbio:2008zf,Hietanen:2008mr,Bursa:2009we,DelDebbio:2010hx}, and SU(3) with 2 sextet (2S) fermions \cite{Shamir:2008pb,Fodor:2009ar,DeGrand:2009hu,Kogut:2010cz,DeGrand:2010na}. These theories
are part of the larger family of theories whose properties are shown
in Figs.~\ref{fig12} and~\ref{fig22}. On each 
of these plots we show $N_f^{\rm{I}}$, as well as three curves related to the 
lower boundary of the conformal window: the curve $N_f^{\rm{MC}}$ where metric confinement sets in, 
the curve $N_f^{\rm{AO}}$ mapped out by the vanishing of the AO $\beta$-function with 
$\gamma=2$, and the curve where causal analyticity breaks down.
The first two provide lower bounds for the conformal window, while
the third gives us an estimate of where non-perturbative effects 
must be important.
%\newline
%\newline
%\indent 
We have also displayed in these figures the SD 
predictions for chiral symmetry breaking (in the usual ladder
approximation). Where chiral symmetry breaking occurs will typically be the lower boundary 
of the conformal window and, in any case, will provide a lower bound 
for it. Unfortunately, although time-honoured, such SD estimates
are known to fail in SQCD \cite{Appelquist:1997gq}.

\subsubsection{$SU(2)$ and $SU(3)$ theories with fundamental flavours}

In the left panels of Figs.~\ref{fig12} and~\ref{fig22} we display estimates for the conformal window of
$SU$ and $SO$ theories ($Sp$ being qualitatively the same as $SU$) with fundamental fermions. It shows that the 
metric confinement and causal analyticity criteria almost coincide in 
all cases. With the exception of $SU(2)$ (and $Sp(4)$), causal analyticity
extends to a slightly lower $N_f$ than metric confinement.  So, in contrast to SQCD,
the whole of the conformal window is causal analytic, suggesting
that it represents a perturbative infra-red conformal phase.

For SU(3) this suggests that the conformal window begins
with $N_f = 10$ and for SU(2) with  $N_f = 7$. However, since the limits
are close together it is important to check whether the coupling remains
small at these limits. In Fig.~\ref{max_nf} we plot the maximal value of the complex
2-loop coupling $\max_{Q^2 \in \field{C}} |N_c x(Q^2)|$ for SU(3), as a function of the scaled flavour variable $\Delta N_f\equiv (N_f-N_f^{MC})/(N_f^I-N_f^{MC})$ taking values from 0 to 1 within the conformal window, and indicate with dots the $N_f=10,12,16$ theories. We see that,
as expected  the coupling remains small for $N_f=16$ and increases as $N_f$ is lowered.
In particular, the coupling is rather large at the lower end of the window,
leaving room for a significant shift, either way, in our estimate
of what is the true region of causal analyticity.

Inside the conformal window the coupling does not decrease linearly with $N_f$ but rather 
increases rapidly as $N_f^{\rm{MC}}$  is approached. This behaviour is plotted in Fig.~\ref{max_nf}.
Although in $SU(3)$
the coupling rapidly increases below $N_f=10$ it should be noted that the coupling is already 
somewhat large by this point.

The so-called 1-family models of technicolour are based on an $SU(2)$ gauge theory with $N_f=8$ in the fundamental representation, see e.g \cite{Farhi:1979zx}.
This theory is well above the
bound on $N_f^{\rm{MC}}$ that follows from metric confinement and within the window of causal analyticity with a relatively small coupling shown in Fig.~\ref{max_nf}, suggesting that the theory is conformal and weakly coupled. 
\subsubsection{Two flavor $SU(2)$ adjoint theories}

The Minimal Walking technicolour (MWT) model \cite{Sannino:2004qp,Gudnason:2006ug} is based 
on $SU(2)$ gauge theory with $N_f=2$ in the adjoint  representation.
Current lattice simulations of this theory suggests that it is conformal 
\cite{Catterall:2007yx,Hietanen:2008mr,DelDebbio:2008zf,Bursa:2009we} 
with a relatively small anomalous mass dimension, close to 
the 1-loop estimate \cite{Bursa:2009we}.

We display in the centre panels of Figs.~\ref{fig12} and~\ref{fig22} what happens for gauge theories with
adjoint fermions. We do so for various values of $N_c$, and for $SO$ 
as well as the $SU$ groups that lattice
calculations have so far focused upon. Results for $Sp$ are identical to those of $SU$. We note that the results
look similar for the $SU$ and $SO$ groups and that
there is no dependence on $N_c$ for a fixed number of adjoint fermions.
This is no surprise, since all our predictions involve some aspect of
the perturbative running. Finally, and most interestingly,
we see from  Fig.~\ref{fig12} that $N_f=2$ is  well above the
bound on $N_f^{\rm{MC}}$ that follows from metric confinement and also
well within the window of causal analyticity. (Which here coincides
with $N_f^{\rm{AO}}$, the $\gamma =2$ bound from the AO $\beta$-function.)  
This strongly suggests that the $N_f=2$ theory is conformal. 

One might be perturbed by the fact that, as we see in Fig.~\ref{fig12},  
causal analyticity extends into the region where metric 
confinement already holds. However, the gap between the two curves is 
small and is presumably consistent with the uncertainty that higher 
order corrections would bring to the location of the breakdown of 
causal analyticity. Following on from the fundamental case, we calculate 
the value of $x$ over the whole complex $Q^2$ plane, so as to see if it is 
everywhere `small' and that our 2-loop analysis can be trusted or if it is 
somewhere `large', increasing the uncertainty in our analysis.

\begin{figure}[htp!]
{\includegraphics[width=\linewidth]{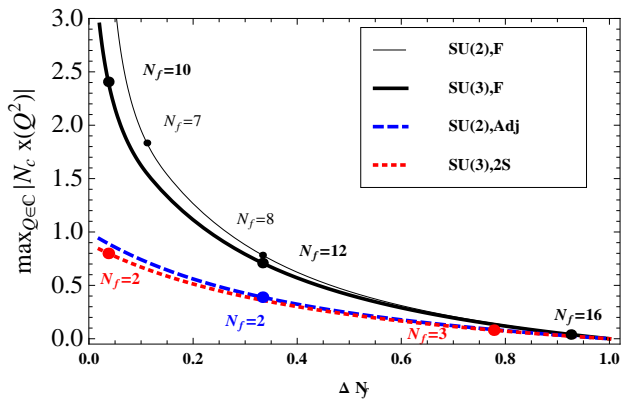}}
\caption{The maximal value of the 2-loop coupling $|N_c x(Q)|$ in the complex 
plane $Q\in \field{C}$, excluding the negative real axis, with $\Delta N_f\equiv (N_f-N_f^{MC})/(N_f^I-N_f^{MC})$ taking values from 0 to 1 within the conformal window for the gauge groups and representations indicated. The location of the theories of Fig.~\ref{normxQ} are indicated in dots.
}
\label{max_nf}
\end{figure}

\begin{figure}[htp!]
{\includegraphics[width=\linewidth]{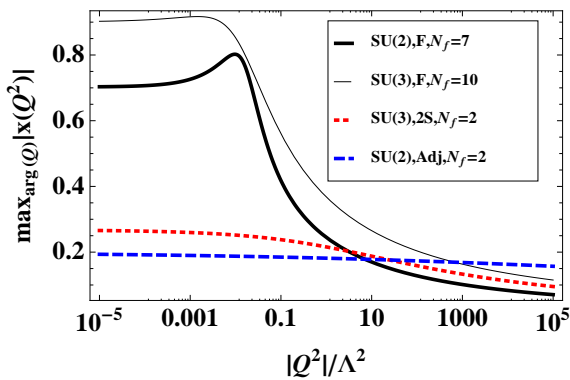}}
\caption{The maximal value of the 2-loop coupling $|x(Q)|$ in the complex $Q\in \field{C}$
plane, excluding the negative real axis, for the theories indicated. A maximum away from $|Q^2|=0$ indicates that the theory is  close to the limit of causal analyticity.}
\label{normxQ}
\end{figure}

The result $\max_{arg(Q^2)} |x(Q^2)|$ for the maximum value of $|x|$ at fixed  $|Q^2|$ for the interesting case of $N_f=2$ is shown in
 Fig.~\ref{normxQ} and $\max_{Q^2 \in \field{C}} |N_c x(Q^2)|$ for general $N_f$ in Fig.~\ref{max_nf}. 
%We show results for the interesting case of $N_f=2$ as well as 
% $N_f=2.5$, which
%is near the upper boundary of the conformal window, and  $N_f=1.5$,
%which is close to where analyticity is lost. 
We observe that,
while the maximum value of  $|x(Q)|$ for $N_f=2$ is not as small as it 
is near the $N_f^I=2.75$ LOAF limit, it is certainly small compared to 
its value at the point near which causality is lost, $N_f^{CA}=1.38$. 
This gives us confidence that at $N_f=2$ the theory really is causally 
analytic and that it is in a perturbative (infra-red) conformal phase. It is thus
consistent with the observation \cite{Bursa:2009we} that $\gamma$ is close to the one-loop prediction.

On the other hand, at $N_f=1.5$ the value of  $|x(Q)|$ is large enough
that it is entirely plausible that a higher order calculation could
shift the loss of analyticity from just below that value of $N_f$ to
above it, so ensuring that metric confinement does not take place within the
region of causal analyticity.


\begin{table}[h]
\begin{center}
%\begin{minipage}{3in}
\begin{tabular}{c||ccccc}
$G$  & $R$& $N_f^{CA}$&$N_f^{MC}$ & $N_f^I$  \\
  \hline \hline \\
$SU(2)$& $F$ & $6.60$ & $6.5$ & $11$     \\
&&&\\
& $Adj$ & $1.38$ & $1.63$ & $2.75$     \\
&&&\\
$SU(3)$& $F$ & $9.68$ & $9.75$ & $16.5$   \\
 &&&\\
& $2S$ & $1.61$ & $1.95$ & $3.3$     \\
\end{tabular}
%\end{minipage}
\end{center}
\caption{The $N_f$ values for loss of causal analyticity $N_f^{CA}$, the lower boundary of the conformal window from metric confinement $N_f^{MC}$, and loss of asymptotic freedom $N_f^I$ for theories considered in the text.}
\label{numbers}
\end{table}

\subsubsection{Two flavor $SU(3)$ sextet theory}

The Next to Minimal Walking technicolour (NMWT) model \cite{Sannino:2004qp,Belyaev:2008yj} 
is based on an $SU(3)$ gauge theory with $N_f=2$ in the two-index symmetric (sextet) representation. Current lattice simulations of 
this theory suggests that it is conformal or near-conformal 
\cite{Fodor:2009ar,Kogut:2010cz,DeGrand:2010na} 
and that it has relatively small anomalous mass dimension, close to 
the 1-loop estimate \cite{DeGrand:2010na}. 

We show in the right panels of Figs.~\ref{fig12} and~\ref{fig22} what happens for $SU$ and $SO$ gauge theories with
fermions in the two-index symmetric representation at
various values of $N_c$ (The symmetric 
representation of $Sp$ is identical to the adjoint of $Sp$).  We note that there is a significant
dependence on $N_c$ and that once again metric confinement sets in within the analyticity
window. However, in contrast to the $SU(2)$ case with adjoint fermions,
metric confinement sets in very close to $N_f=2$. (See table~\ref{numbers}). Thus we expect
that the $N_f=2$ theory is very close to the lower boundary of the
conformal window.

Once again we compute the value of $|x(Q)|$ from the 2-loop 
$\beta$-function in the whole of the $Q^2$ complex-plane, but this
time for SU(3) with 2 sextet fermions. The result for
$\max_{arg(Q^2)} |x(Q^2)|$ is shown 
in Fig.~\ref{normxQ} for $N_f = 2$ and $\max_{Q^2 \in \field{C}} |N_c x(Q^2)|$ for general $N_f$ in Fig.~\ref{max_nf}, where we also indicate $N_f=3$ which is near the upper boundary of the conformal window.
%, and  $N_f=1.65$,
%which is just about where analyticity is lost. 
We observe that
the maximum value of $|N_c x|$ for $N_f=2$ is relatively small, compared to the $SU(3)$ theory with 10 fundamental flavors, although significantly larger than it is in the case of adjoint fermions. The corresponding value of $\alpha_s=\pi x$ is also larger than the value $\alpha_s\sim 0.5$ at which, in QCD, one typically begins to worry about the convergence of perturbation theory, while for MWT the coupling is indeed slightly smaller. (Though, it is not obvious how to compare the size of the couplings across theories with fermions in different representations.) 

This leaves it unclear whether, at the point at which metric confinement
sets in and conformality is lost, the theory is still consistently 
perturbative.

\subsection{Conclusions}

In this letter we have discussed the implications of `metric confinement'
and `causal analyticity' for theories that are being actively studied
using lattice techniques in the search for walking near-conformal
field theories. 

We noted that in the case of SQCD, where Seiberg duality
gives us a precise description of the conformal window, both these
criteria work very well: metric confinement predicts the precise
location of the lower boundary of that window while causal analyticity
predicts that the theory becomes strongly coupled in the lower part of
the window, as required by the weak-strong duality. On the other
hand, the widely used SD calculations for where chiral
symmetry breaking sets in, are very badly off in SQCD. This is part
of our motivation for bringing these other criteria into play.

It is interesting that for the theories considered here, generically 
perturbation theory is consistent all the way down to the lower end of their conformal 
window as determined by metric confinement, and so the mass anomalous
dimension at the fixed point can be plausibly estimated in 1-loop 
perturbation theory. Doing so we find $\gamma(x_{\FP})=0.6, \  1.34$
for the MWT and NMWT theories respectively. Going to the next order in $\overline{MS}$ the values of $\gamma$ change by about $10 \%$ while the corresponding 
predictions from the AO $\beta$-function, setting $\beta(x_{\FP})=0$ in Eq.~\ref{AOBF} 
are $\gamma(x_{\FP})= 0.75 , \ 1.3$.  This can be
compared to the results of lattice simulations 
\cite{Bursa:2009we,DeGrand:2009hu,DeGrand:2010na,DelDebbio:2010hx} which suggest anomalous 
dimensions consistent with the 1-loop result, albeit with the caveat
that for the MWT model the simulations find a fixed point which 
is a factor  two smaller than the two-loop result we have used.

In the case of MWT both criteria suggest that this theory lies 
well within a perturbative infra-red conformal phase. 
By contrast, NMWT appears to be almost on the boundary
of the lower conformal window. This is certainly consistent with
the mixed messages one has been getting from different lattice 
calculations on this theory \cite{DeGrand:2009hu,Kogut:2010cz,DeGrand:2010na}.
The possibility that this theory lies just outside the conformal window, which is possible 
because, strictly speaking, metric confinement provides a lower bound 
on where confinement sets in, makes it an interesting candidate walking technicolour model in itself. For example,  the presence of four fermion operators, arising from extended technicolour interactions, can modify the conformal window and anomalous dimensions (indeed it can do so in all the theories we consider here\cite{Yamawaki:1996vr}).

As already observed in  \cite{Gardi:1998ch}, metric confinement suggests 
that the conformal window for $SU(3)$ with $N_f$ fundamental fermions 
begins at $N_f=10$, as we can infer from Fig.~\ref{fig12}. 
As pointed out in \cite{Gardi:1998ch} causal analyticity
extends just below $N_f=10$, suggesting that the whole conformal
window is weakly coupled. However if one actually looks at the
coupling $x$ in the $N_f=10$ theory, one finds that its value is quite
large, as shown in  Fig.~\ref{normxQ}. So if it turns out that the  $N_f=10$ 
theory does not, in fact, lie in the conformal window
%, which is possible 
%because, strictly speaking, metric confinement provides a lower bound 
%on where confinement sets in, 
then again this opens the possibility of the kind 
of large anomalous dimension that walking phenomenology needs.  
On the other hand, there appears to be little doubt that the $N_f=12$
theory does lie well inside the conformal window, and  $N_f=9$ 
well outside.

Very similar remarks apply to  $SU(2)$ with $N_f$ fundamental fermions. The conformal window should begin at $N_f=7$, which is similar to $N_f=10$ in $SU(3)$. $N_f=8$ is very similar to $N_f=12$ in $SU(3)$, while
$N_f=6$ lies just inside the region of metric confinement, albeit
still in the region of causal analyticity.

\section*{Acknowledgements}

MTF thanks E. Gardi for the initial discussion that led to this study and acknowledges
a Villum Kann Rasmussen Foundation Fellowship. We thank D. D. Dietrich, F. Sannino and R. Zwicky for useful discussions.

\end{document}